\documentclass[a4paper,11pt]{article}

\usepackage{graphicx} 
\usepackage[margin=1in]{geometry}
\usepackage{algorithm}
\usepackage[noend]{algpseudocode}
\usepackage{amsthm}
\usepackage{amssymb,amsmath}  
\usepackage{thmtools}  
\usepackage{mathtools}
\usepackage[colorlinks=true, allcolors=blue]{hyperref}
\usepackage[capitalize]{cleveref}
\usepackage{placeins}
\usepackage{subcaption}
\usepackage{booktabs}

\usepackage{newtxtext}
\usepackage{newtxmath}

\newcommand{\defDSproblem}[3]{
\vspace{2mm}
\noindent\fbox{
   \begin{minipage}{0.9\columnwidth}
   \textsc{#1}\\
   {\bf{Preprocess:}} #2  \\
   {\bf{Query:}} #3
   \end{minipage}
   }
   \vspace{2mm}
}

\usepackage{enumitem}
\setlist[1]{labelindent=\parindent}
\setlist[enumerate]{label=(\arabic*)}
\setlist[itemize]{noitemsep}
\usepackage{orcidlink}
\usepackage{authblk}

\author[1]{Solon P. Pissis \orcidlink{0000-0002-1445-1932}}
\author[2]{Sharma V. Thankachan \orcidlink{0000-0002-6852-1035}\thanks{Supported in part by U.S. National Science Foundation (NSF) Awards CCF-2316691 and CCF-2315822.}}
\affil[1]{The Cyprus Institute, Nicosia, Cyprus}
\affil[2]{North Carolina State University, Raleigh, NC, USA}
\date{\today}

\declaretheorem[numberwithin=section]{theorem}

\declaretheorem[sibling=theorem]{lemma}

\declaretheorem[sibling=theorem]{proposition}



\usepackage{xspace}
\usepackage{tikz}
\usetikzlibrary{decorations.pathreplacing}

\def\dd{\mathinner{.\,.}}

\newcommand{\hpl}{\textsf{hp-leaf}}

\newcommand{\str}{\textsf{str}}
\newcommand{\size}{\textsf{size}}
\newcommand{\cO}{\mathcal{O}}

\newcommand{\ST}{\textsf{ST}}
\newcommand{\locus}{\textsf{locus}}
\newcommand{\SA}{\textsf{SA}}
\newcommand{\heavy}{\textsc{Heavy}\xspace}
\newcommand{\light}{\textsc{Light}\xspace}
\newcommand{\Occ}{\textsf{Occ}}
\newcommand{\Occno}{\Occ^{\text{NO}}}

\newcommand{\Occco}{\Occ^{\beta}}

\newcommand{\parent}{\textsf{parent}}
\newtheorem{problem}{Problem}
\newcommand{\NOC}{NOC\xspace}

\newcommand{\COC}{COC\xspace}
\newcommand{\POC}{POC\xspace}

\title{Space-Efficient Text Indexing for Advanced Counting Queries}

\date{\vspace{-.5cm}}

\begin{document}

\maketitle

\begin{abstract}
In the classical problem of \emph{text indexing}, the objective is to preprocess a text \(T=T[1\dd n]\) of length $n$ over an integer alphabet \([0\dd\sigma)\) into a data structure that supports efficient pattern matching queries. Given an input pattern \(P=P[1\dd m]\) of length $m$, a reporting query lists the set \(\Occ_T(P)\) of occurrences (starting positions) of $P$ in $T$, whereas a counting query returns \(|\Occ_T(P)|\). The suffix tree of \(T\), which requires \(\cO(n)\) words of space, or equivalently \(\cO(n \log n)\) bits, supports reporting queries in \(\cO(m + |\Occ_T(P)|)\) time and counting queries in \(\cO(m)\) time. Space-efficient suffix-tree variants use only \(\cO(n \log \sigma)\) bits of space and support reporting queries in \(\cO(m + |\Occ_T(P)| \cdot \log_{\sigma}^{\epsilon} n)\) time, where \(\epsilon>0\) is an arbitrarily small constant, and counting queries in \(\cO(m)\) time. Notably, space-efficient text indexes for more \emph{advanced counting queries} have largely resisted achieving the same optimal (or near-optimal) query times when using less than \(\Theta(n \log n)\) bits of space.

In this paper, we change this situation by proposing a \emph{general framework} that supports advanced counting queries, denoted by \(\phi_T(P)\), provided that two mild conditions are satisfied on \(\phi_T\): (1)~\(\phi_T(P)\) is an integer in \([0\dd|\Occ_T(P)|]\) that can be computed from \(\Occ_T(P)\) via sorting; and (2) for every prefix \(Q\) of \(P\), we have \(\phi_T(Q) \geq \phi_T(P)\). Clearly (1) and (2) hold for \(\phi_T(P)\coloneqq|\Occ_T(P)|\). We show that our framework supports \emph{optimal} $\cO(m)$-time queries using $\cO(n \log \sigma + n(\log \log n)^2)$ bits or \emph{near-optimal} $\cO(m\log_{\sigma}^{\epsilon} n)$-time queries using $\cO(n \log \sigma)$ bits. As an example, our trade-offs improve upon: the classical MAST index for counting non-overlapping occurrences [Apostolico and Preparata, Algorithmica, 1996; Brodal et al., ICALP 2002] from $\cO(n \log n)$ bits and $\cO(m)$-time queries to $\cO(n \log \sigma + n(\log \log n)^2)$ bits and $\cO(m)$-time queries; the recently introduced compact data structure for counting non-overlapping occurrences [Bals et al., ESA 2026] from $\cO(n \log \sigma)$ bits and $\cO(m\log n\log_{\sigma}^{\epsilon} n)$-time queries to $\cO(n \log \sigma)$ bits and $\cO(m\log_{\sigma}^{\epsilon} n)$-time queries.
\end{abstract}

\thispagestyle{empty}
\clearpage
\pagenumbering{arabic}
\setcounter{page}{1}

\section{Introduction}

Text indexing is a fundamental problem in computer science with broad applications,
notably in computational biology~\cite{DBLP:books/cu/Gusfield1997} and information retrieval~\cite{DBLP:books/aw/Baeza-YatesR2011}. The objective is to preprocess a string $T=T[1\dd n]$ of length $n$ over an integer alphabet $[0\dd\sigma)$ into a data structure, referred to as the \emph{index}, that efficiently answers queries for a string $P=P[1\dd m]$ of length $m \leq n$. In this context, we use the term \emph{text} for string $T$ and the term \emph{pattern} for string $P$.
The two most common query types supported by text indexes are \emph{reporting} and \emph{counting}: reporting queries return the set $\Occ_T(P)$ of all starting positions (occurrences) of $P$ in $T$, whereas counting queries return the cardinality $|\Occ_T(P)|$.

Research on text indexing began in the early 1970s with Weiner's introduction of suffix trees~\cite{DBLP:conf/focs/Weiner73}. This seminal data structure requires $\cO(n)$ words of space, or equivalently \(\cO(n \log n)\) bits, can be constructed in $\cO(n)$ time~\cite{DBLP:conf/focs/Farach97}, and supports reporting queries in $\cO(m + |\Occ_T(P)|)$ time and counting queries in $\cO(m)$ time. Later efforts in text indexing concentrated primarily on reducing space usage---the principal practical bottleneck---frequently accepting slower query times as a trade-off. Major advances along this line include the suffix array~\cite{DBLP:books/daglib/GonnetBS92,DBLP:journals/siamcomp/ManberM93}, the compressed suffix array~\cite{DBLP:journals/siamcomp/GrossiV05}, the FM-index~\cite{DBLP:journals/jacm/FerraginaM05}, the $r$-index~\cite{DBLP:conf/soda/GagieNP18}, and other indexes designed for highly repetitive text collections~\cite{DBLP:journals/csur/Navarro21}. This line of research has culminated in space-efficient suffix-tree variants that occupy only $\cO(n \log \sigma)$ bits of space and support reporting queries in $\cO(m + |\Occ_T(P)| \cdot \log_{\sigma}^{\epsilon} n)$ time~\cite{DBLP:journals/siamcomp/GrossiV05}, where $\epsilon>0$ is an arbitrarily small constant, while maintaining $\cO(m)$ time for counting queries~\cite{DBLP:journals/talg/BelazzouguiN14}.

In this paper, we are concerned with more \emph{advanced counting queries} arising in text indexing.
Suppose, as an example, that one wants to preprocess a text $T$ to answer queries of the following type for a given pattern $P$: count the largest set of \emph{non-overlapping occurrences} of $P$ in $T$~\cite{DBLP:journals/algorithmica/ApostolicoP96,DBLP:conf/icalp/BrodalLOP02}; or count the number of \emph{consecutive occurrences} of $P$ in $T$ that are at distance at most $\beta$~\cite{DBLP:journals/tcs/NavarroT16,DBLP:conf/esa/Bals26}.
Such, more advanced, types of queries typically admit \(\cO(n \log n)\)-bit data structures with $\cO(m)$-time queries. Unfortunately, \emph{space-efficient} data structures for such counting queries have largely resisted achieving the same optimal (or even near-optimal) query times using less than \(\Theta(n \log n)\) bits of space. 

This brings us to the following basic question:

\begin{center}
    \fcolorbox{gray!50}{gray!15}{%
        \begin{minipage}{0.9\linewidth}
            \centering
            \emph{Can we break through the \(\Theta(n \log n)\)-bit $\cO(m)$-time trade-off \\for advanced counting queries in text indexing?}
        \end{minipage}%
    }
\end{center}

\subsection{Our Contributions}

We essentially settle the above question in the affirmative by proposing a \emph{general framework} that supports advanced counting queries, provided that two mild conditions are satisfied. Let us make this precise.

\begin{problem}\label{prob:main}
Preprocess a text \(T=T[1\dd n] \in [0\dd\sigma)^n\), so that given a pattern \(P=P[1\dd m]\) as a query, we can output \(\phi_T(P)\), where $\phi$ is a function of \(P\) (and \(T\)) satisfying the following two conditions:
\begin{description}
    \item[1. Computing \(\phi_T\):] \(\phi_T(P)\) is an integer in \([0\dd |\Occ_T(P)|]\) that can be computed from the set \(\Occ_T(P)\) of occurrences of \(P\) in \(T\) in \(\cO(m+|\Occ_T(P)| \log\log n)\) time.   
    \item[2. Monotonicity:] If \(Q\) is a prefix of \(P\), then \(\phi_T(P) \leq \phi_T(Q)\). 
\end{description}
\end{problem}

We next provide intuition for the above two conditions and argue why they are mild (and somewhat even standard). The first condition (computing \(\phi_T\)) simply requires that the answer may be derived from the set \(\Occ_T(P)\) of occurrences of $P$ in $T$ using integer sorting. Indeed, sorting $n$ positive integers can be done in \(\cO(n \log\log n)\) time~\cite{DBLP:journals/jal/Han04}. This  condition is satisfied by many natural counting problems in text indexing. For example, for counting non-overlapping occurrences, one can sort \(\Occ_T(P)\) and greedily select occurrences via a left-to-right scan. The second condition (monotonicity) requires that extending a pattern by one letter cannot increase the count---a natural constraint that holds for many counting problems arising in text indexing. For example, for consecutive occurrences with gap at most \(\beta\), extending a pattern \(Q\) to \(P\) by one letter can only reduce occurrences, and each remaining consecutive pair in \(P\) corresponds to at least one consecutive pair in \(Q\) with gap at most \(\beta\), ensuring that \(\phi_T(P) \leq \phi_T(Q)\).

Our main result is given below---throughout this paper, $\epsilon>0$ denotes an arbitrarily small constant.

\begin{theorem}[Main Result]\label{thm:main}
For \Cref{prob:main}, the following space-time trade-offs are possible:
    \begin{enumerate}
    \item An \(\cO(n \log \sigma + n(\log \log n)^2)\)-bit data structure with \(\cO(m)\) query time.

    \item An \(\cO(n \log \sigma)\)-bit data structure with \(\cO(m \log_{\sigma}^{\epsilon} n)\) query time.
\end{enumerate}
\end{theorem}

\paragraph{Applications.} We choose here to showcase three headline applications of \Cref{thm:main} that improve upon the state-of-the-art indexes; however, we anticipate that \Cref{thm:main} has many more applications.

\begin{description}
    \item[Non-Overlapping Occurrences:] Text indexing for reporting or counting non-overlapping occurrences is a well-studied problem in combinatorial pattern matching~\cite{DBLP:journals/algorithmica/ApostolicoP96,DBLP:conf/icalp/BrodalLOP02,DBLP:conf/isaac/CohenP09,DBLP:journals/algorithmica/GangulyST20,DBLP:journals/tcs/HooshmandAKT21,DBLP:journals/tcs/GibneyMT25,DBLP:conf/esa/Bals26}. Counting the largest set of non-overlapping occurrences of a pattern $P$ in text $T$ can be done using the classical minimal augmented suffix tree (MAST) of Apostolico and Preparata~\cite{DBLP:journals/algorithmica/ApostolicoP96} (see also~\cite{DBLP:conf/icalp/BrodalLOP02}). It uses $\cO(n \log n)$ bits and achieves $\cO(m)$ query time. The first trade-off of \Cref{thm:main} improves this to $\cO(n \log \sigma + n(\log \log n)^2)$ bits while maintaining $\cO(m)$ query time, thus breaking the $\Theta(n \log n)$-bit barrier for this problem whenever $\log\sigma=o(\log n)$. Very recently, Bals, Charalampopoulos, Lachish, Pissis, and Verbeek~\cite{DBLP:conf/esa/Bals26} showed an index 
    that uses $\cO(n\log \sigma)$ bits and achieves $\cO(m\log n \log_{\sigma}^{\epsilon} n)$ query time. The second trade-off of \Cref{thm:main} improves the query time to $\cO(m\log_{\sigma}^{\epsilon} n)$ while maintaining $\cO(n\log \sigma)$ bits.

\item[Consecutive Occurrences:] Text indexing for reporting or counting consecutive occurrences is also a well-studied problem. In particular, many variants of consecutive occurrences have been studied in the literature~\cite{DBLP:journals/algorithmica/BilleG14,
    DBLP:conf/stacs/BilleGLPRS24,
    DBLP:journals/tcs/BilleGPRS22,
    DBLP:journals/algorithmica/BilleGPS23,
    DBLP:journals/mst/BilleGVV14,
    DBLP:conf/cpm/BrodalLPS99,
    DBLP:journals/is/GawrychowskiGSS26,
    DBLP:journals/algorithmica/IliopoulosR09,
    DBLP:conf/wads/KellerKL07,
    DBLP:conf/soda/Muthukrishnan02,
    DBLP:journals/tcs/NavarroT16,
    DBLP:conf/esa/Radoszewski23,DBLP:conf/esa/Bals26}. For the most basic version of counting consecutive occurrences of $P$ in $T$ with gap at most $\beta$, the recent framework of Bals et al.~\cite{DBLP:conf/esa/Bals26} yields an $\cO(n \log n)$-bit data structure with $\cO(m)$ query time. 
    The first trade-off of \Cref{thm:main} improves
    the space to $\cO(n \log \sigma + n(\log \log n)^2)$ bits with the same $\cO(m)$ query time.

\item[Property-Restricted Occurrences:] In this problem, in addition to $T$, we are given a hereditary property $\Pi$ of intervals (closed under subintervals) from $[1\dd n]$, represented by an array $\pi$, where $\pi[i]$ is the rightmost endpoint of a valid interval starting at $i$. The objective is to report or count occurrences of a pattern $P$ that start at positions $i$ such that the interval $[i\dd i+m-1]$ satisfies $\Pi$, i.e., $i+m-1 \leq \pi[i]$. This is a well-studied problem~\cite{DBLP:journals/tcs/AmirCIKZ08,DBLP:journals/ipl/IliopoulosR08,DBLP:journals/ipl/JuanLW09,DBLP:journals/iandc/HonPST13,DBLP:journals/iandc/BartonK0PR20,DBLP:journals/jea/Charalampopoulos20} mainly due to its direct applications in indexing weighted sequences~\cite{DBLP:journals/tcs/AmirCIKZ08,DBLP:conf/edbt/BiswasPTS16,DBLP:conf/icde/Gabory0LPZ24}.
For counting occurrences of $P$ that satisfy $\Pi$, the best-known solution uses $\cO(n \log n)$ bits and achieves $\cO(m)$ query time~\cite{DBLP:journals/iandc/BartonK0PR20}. The first trade-off of \Cref{thm:main} improves the space to $\cO(n \log \sigma + n(\log \log n)^2)$ bits while preserving $\cO(m)$ query time.
\end{description}

\paragraph{Our Techniques.}
Bals et al.~\cite{DBLP:conf/esa/Bals26} have recently shown how text-indexing reporting data structures can be used to efficiently support counting queries. The authors proved that it suffices to store the counts for at most $n$ ``frequent'' substrings of $T$ in a preprocessing step, and then use a reporting query to count the ``infrequent'' substrings. Their framework typically achieves optimal queries using $\cO(n\log n)$ bits or near-optimal queries using $\cO(n\log \sigma)$ bits. Here, we refine the strategy of Bals et al.~as follows. 

For ``very frequent'' substrings, we store the answers using $\cO(n)$ bits. For ``frequent'' substrings, we store the answers using $\cO(n(\log \log n)^2)$ bits. For ``infrequent'' substrings, we also resort to reporting. The main ingredients to realize this decomposition effectively are Elias--Fano encoding~\cite{DBLP:journals/jacm/Elias74,DBLP:reports/MITCS/Fano71}, heavy-path decomposition~\cite{DBLP:journals/jcss/SleatorT83}, and succinct data structures~\cite{DBLP:conf/focs/Jacobson89}. This yields our first trade-off: an $\cO(n \log \sigma + n(\log \log n)^2)$-bit data structure with $\cO(m)$ query time. By balancing the space allocated to stored answers for frequent substrings against the cost of
reporting the occurrences of infrequent substrings, we obtain our second trade-off: an $\cO(n \log \sigma)$-bit data structure with $\cO(m \log_{\sigma}^{\epsilon} n)$ query time. Given the broad consequences of our main result, the presented solution is surprisingly simple, as it spans about two pages, after developing our encoding toolbox, which is also elementary.

\paragraph{Paper Organization.} In \Cref{sec:prel}, we present the necessary preliminaries.
In \Cref{sec:toolbox}, we develop our encoding toolbox.
In \Cref{sec:optimal_time}, we present the solution for the first trade-off.
In \Cref{sec:optimal_space}, we present the solution for the second trade-off.
In \Cref{sec:apps}, we present the headline applications.

\section{Preliminaries}\label{sec:prel}

A \emph{string} is a finite sequence of characters over an alphabet $\Sigma$.
Depending on the context, we may use \emph{text} or \emph{pattern} as synonyms of the term string.
In this work, we consider strings over an integer alphabet $\Sigma=[0\dd\sigma)$, $\sigma\geq 2$.
We denote the length of a string $T$ by $|T|$.
For any index $i \in [1\dd |T|]$, we denote the $i$-th character of $T$ by $T[i]$.
A string of the form $T[i\dd j] = T[i]T[i+1]\cdots T[j]$,
where $1 \le i \le j \le |T|$, is called a \emph{fragment} of $T$.
If $T$ is known, we can represent any fragment $T[i\dd j]$ in $\cO(1)$ space as the pair $(i,j)$.
A fragment of the form $T[1\dd j]$ is called a \emph{prefix} of $T$, and a fragment of the form $T[i\dd |T|]$ is called a \emph{suffix} of $T$.
The \emph{concatenation} of two strings $S$ and $T$ is denoted by $ST$ or $S \cdot T$.

The \emph{suffix tree} of text \(T=T[1\dd n] \in [0\dd\sigma)^n\), denoted by $\ST_T$, is the compacted trie of all suffixes of \(T\)~\cite{DBLP:conf/focs/Weiner73}. 
For convenience, we will always assume that $T$ ends with a sentinel symbol $\$\notin\Sigma$. In this case, $\ST_T$ has exactly \(n\) leaves and at most \((n-1)\) internal nodes. 
Let $\ell_1, \ell_2, \dots, \ell_n$ denote the leaves of $\ST_T$ in the left-to-right order (e.g., the leaves output in a depth-first traversal). Each edge of $\ST_T$ is labeled implicitly by a fragment $T[i\dd j]$ using a pair $(i,j)$ of indices.
For any node \(u\) of $\ST_T$, let \(\str(u)\) denote the string obtained by concatenating the edge labels along the path from the root to \(u\). We define \(\size(u)\) as the number of leaves in the subtree rooted at \(u\). Given a pattern \(P\), its \emph{locus}, denoted by \(\locus(P)\) (if it exists), is the highest node $u$ whose \(\str(u)\) has \(P\) as a prefix. The interval of leaf indices in the subtree rooted at \(\locus(P)\), denoted by \([s\dd e]\), is called the \emph{suffix range} of \(P\); i.e., these are leaves $\ell_s,\ell_{s+1},\ldots,\ell_e$.
The \emph{suffix array} of \(T\) is an array \(\SA_T[1\dd n]\), where \(\SA_T[i]\) is the starting position of the \(i\)-th lexicographically smallest suffix of \(T\)~\cite{DBLP:journals/siamcomp/ManberM93}. All these data structures require \(\Theta(n\log n)\) bits of space. The set of \emph{occurrences} (starting positions) of $P$ in $T$ is denoted by \(\Occ_T(P)\).
Given a pattern \(P\), we can locate \(\locus(P)\) in $\ST_T$, and thus obtain its suffix range \([s\dd e]\) and the number of occurrences \(|\Occ_T(P)|=e-s+1\), in the optimal \(\cO(|P|)\) time~\cite{DBLP:conf/focs/Weiner73}. All occurrences of \(P\), namely the entries \(\SA_T[s\dd e]\), can then be reported in an additional \(\cO(|\Occ_T(P)|)\) time.

The \emph{compressed suffix array} of Grossi and Vitter~\cite{DBLP:journals/siamcomp/GrossiV05} provides an $\cO(n \log \sigma)$-bit encoding of the suffix array of $T$ that supports $\SA_T[i]$ queries, for any $i\in[1\dd n]$, in $\cO(\log_{\sigma}^{\epsilon} n)$ time. Sadakane showed that, by using $\cO(n)$ additional bits, we can obtain an $\cO(n \log \sigma)$-bit encoding of $\ST_T$  supporting a repertoire of navigational operations, including lowest common ancestor (LCA) queries~\cite{DBLP:journals/mst/Sadakane07}. 
Very recently, Thankachan showed that $\ST_T$ can in fact be encoded using only $\cO(n\log\log\log \sigma)$ bits of space, auxiliary to the text $T$, while supporting navigational operations in polylogarithmic time~\cite{DBLP:journals/corr/abs-2607-17287}.
Belazzougui and Navarro~\cite{DBLP:journals/talg/BelazzouguiN14} proposed a refined version of the FM-index~\cite{DBLP:journals/jacm/FerraginaM05}  of size $n \log \sigma + o(n \log \sigma) + \cO(n)$ bits that supports computing the suffix range of a pattern $P$ in $\cO(|P|)$ time.
Therefore, using these structures within a total space of $\cO(n \log \sigma)$ bits, we can achieve the following. Given a pattern $P$, we can compute the suffix range \([s\dd e]\) of $P$ in $\cO(|P|)$ time. Then, \(\locus(P)\), which is the LCA of $\ell_s$ and $\ell_e$ in $\ST_T$,  can be computed in $\cO(1)$  time via querying the encoding of $\ST_T$ topology.
Subsequently, all occurrences of $P$ in $T$ can be reported in additional $\cO(|\Occ_T(P)|  \log_{\sigma}^{\epsilon} n)$ time.

We assume the standard word-RAM model with words of $\Theta(\log n)$ bits, where $n$ is the input size.

\section{Our Encoding Toolbox}\label{sec:toolbox}

\subsection{Elias--Fano Encoding}

Let \(x_1,x_2,\ldots,x_m\) be a sequence of $m$ integers such that \(0\leq x_i<u\), for some integer $u>0$. If the sequence is strictly increasing (or strictly decreasing), the Elias--Fano (EF) encoding~\cite{DBLP:journals/jacm/Elias74,DBLP:reports/MITCS/Fano71} represents it using \(m\log(u/m)+\cO(m)\) bits of space. More generally, if the sequence is non-decreasing (or non-increasing), the generalized EF encoding (cf.~\cite{DBLP:conf/wsdm/Vigna13}) achieves the same asymptotic space bound. In both cases, the encoding supports retrieving \(x_i\), for any \(i\in[1\dd m]\), in \(\cO(1)\) time.

We now prove some useful lemmas. 

\begin{lemma}\label{lem:EF_small_domain}
Let \(f:[1\dd d]\rightarrow[1\dd r]\) be a non-decreasing (or non-increasing) function, where \(d\le r\). Then \(f\) can be represented using \(\cO(d\log(r/d)+d)\) bits while retrieving \(f(i)\), for any \(i\in[1\dd d]\), in \(\cO(1)\) time.
\end{lemma}

\begin{proof}
The sequence \(f(1),f(2),\ldots,f(d)\) is non-decreasing (or non-increasing) over the universe \([1\dd r]\). Therefore, by the generalized EF encoding, it can be represented using
\(d\log(r/d)+\cO(d)=\cO(d\log(r/d)+d)\) bits while supporting the retrieval of \(f(i)\), for any \(i\in[1\dd d]\), in \(\cO(1)\) time.
\end{proof}

We next present \Cref{lem:EF_small_range}, whose proof relies on a somewhat novel application of \Cref{lem:EF_small_domain}.

\begin{lemma}\label{lem:EF_small_range}
Let \(f:[1\dd d]\rightarrow[1\dd r]\) be a non-decreasing (or non-increasing) function, where \(d>r\). Then \(f\) can be represented using \(\cO(r\log(d/r)+r)\) bits while retrieving \(f(i)\), for any \(i\in[1\dd d]\), in \(\cO(i)\) time.
\end{lemma}

\begin{proof}
Partition \([1\dd d]\) into the maximal intervals on which \(f\) is constant. Namely, we partition 
\([1\dd d]\) into \([1\dd d_1],[d_1+1\dd d_2],\ldots,[d_{z-1}+1\dd d_z]\), such that all values within the same interval map to the same function value, and distinct intervals correspond to distinct function values. Since the range of \(f\) is contained in \([1\dd r]\), we have \(z\le r\).
We store the pairs
\[
(d_1,f(d_1)),(d_2,f(d_2)),\ldots,(d_z,f(d_z)).
\]
By construction, the sequence \(d_1,d_2,\ldots,d_z\) is strictly increasing over the universe \([1\dd d]\), and the sequence \(f(d_1),f(d_2),\ldots,f(d_z)\) is also strictly monotone. Hence, both sequences can be represented using EF encoding in a total of
\[
\cO(z\log(d/z)+z) \subseteq \cO(r\log(d/r)+r)
\]
bits. Given \(i\in[1\dd d]\), to retrieve \(f(i)\), we linearly scan the stored breakpoints $d_1,d_2,\ldots,d_z$ until finding the first \(d_j\ge i\), and return \(f(d_j)\). This requires \(\cO(j)\) time. By the choice of \(j\), we have \(d_{j-1}<i\), and since every interval is non-empty, \(d_{j-1}\ge j-1\); thus \(j-1<i\). Hence, the query time is \(\cO(i)\).
\end{proof}

We next generalize \Cref{lem:EF_small_domain} and \Cref{lem:EF_small_range} to the scenario where we need to encode a collection of non-decreasing (or non-increasing) functions.\footnote{We present the proofs for the non-decreasing case, but the changes for the non-increasing case are trivial.} A direct approach is to apply the above lemmas to each function separately. However, this may require storing an additional pointer for each encoded function, and the resulting overhead can dominate the space required by the encodings themselves. The following two lemmas provide a joint representation that avoids this additional overhead.

\begin{lemma} \label{lem:EF_compact}
For \(k\in[1\dd h]\), let \(f_k:[1\dd d_k]\rightarrow[1\dd r_k]\) be a non-decreasing (or non-increasing) function, where \(d_k\le r_k\). Let \(d\coloneqq\sum_{k=1}^{h}d_k\) and \(r\coloneqq\sum_{k=1}^{h}r_k\). Then the collection \(\{f_1,f_2,\ldots,f_h\}\) can be represented using \(\cO(d\log(r/d)+d)\) bits while retrieving \(f_k(i)\), for any \(k\in[1\dd h]\) and \(i\in[1\dd d_k]\), in \(\cO(1)\) time.
\end{lemma}
\begin{proof}
We define a single function \(F:[1\dd d]\rightarrow[1\dd r]\) by concatenating the functions \(f_1,\ldots,f_h\) and shifting their ranges. Specifically, for \(k\in[1\dd h]\) and \(i\in[1\dd d_k]\), let \(F(i+\sum_{j=1}^{k-1}d_j)\coloneqq f_k(i)+\sum_{j=1}^{k-1}r_j\). The resulting function \(F\) is non-decreasing. By \Cref{lem:EF_small_domain}, \(F\) can be represented using \(\cO(d\log(r/d)+d)\) bits while supporting queries in \(\cO(1)\) time. We store the domain boundaries \(\sum_{j=1}^{k}d_j\) and the range offsets \(\sum_{j=1}^{k}r_j\), for \(k\in[1\dd h]\), using EF encoding. This requires \(\cO(h\log(d/h)+h\log(r/h)+h)\) bits and supports retrieving the \((k-1)\)-th boundary and the \((k-1)\)-th range offset, for any \(k\in[1\dd h]\), in \(\cO(1)\) time (both \(0\) if \(k=1\)).
Since \(h\le d\) and \(h\le r\), this space is absorbed by the \(\cO(d\log(r/d)+d)\)-bit bound. Given a query specified by the pair $(k, i)$, we first use the stored boundaries to locate the corresponding position in \(F\), then query \(F\), and, finally, subtract the range offset \(\sum_{j=1}^{k-1}r_j\) to recover \(f_k(i)\). Thus, the total query time is \(\cO(1)\).
\end{proof}

\begin{lemma} \label{lem:EF_opt_time}
For \(k\in[1\dd h]\), let \(f_k:[1\dd d_k]\rightarrow[1\dd r_k]\) be a non-decreasing (or non-increasing) function, where \(d_k\geq r_k\). Let \(d\coloneqq\sum_{k=1}^{h}d_k\) and \(r\coloneqq\sum_{k=1}^{h}r_k\). Then the collection \(\{f_1,f_2,\ldots,f_h\}\) can be represented using \(\cO(r\log(d/r)+r)\) bits while retrieving \(f_k(i)\), for any \(k\in[1\dd h]\) and \(i\in[1\dd d_k]\), in \(\cO(i)\) time.
\end{lemma}

\begin{proof}
We simply combine the approaches in \Cref{lem:EF_small_range} and \Cref{lem:EF_compact} to arrive at the result.

For each function \(f_k\), we partition its domain into maximal intervals on which it is constant. Let \(z_k\) denote the number of such intervals, and store the (right) endpoints $e_{k,1},e_{k,2},\dots,e_{k,z_k}=d_k$ of these intervals along with their corresponding function values. Since
each interval has a distinct function value and the range has size
$r_k$, we have $z_k\le r_k$. The total number of stored intervals is \(z=\sum_{k=1}^{h}z_k\le r\).

We combine all interval endpoints into a single monotone sequence over the universe \([1\dd d]\), by adding cumulative domain offsets \(\sum_{j=1}^{k-1}d_j\). We similarly combine the corresponding function values into a single monotone sequence over the universe \([1\dd r]\), by adding cumulative range offsets \(\sum_{j=1}^{k-1}r_j\).  Both sequences have length \(z \le r\) and, since $d\geq r$, they can be represented using EF encoding in \(\cO(z\log(d/z)+z) \subseteq \cO(r\log(d/r)+r)\) bits.

To identify the block of breakpoints corresponding to $f_k$, let
\(Z_k\coloneqq\sum_{j=1}^k z_j\), and store the sequence \(Z_1,Z_2,\ldots,Z_h\) using EF encoding.
This requires \(\cO(h\log(z/h)+h)\) additional bits, which is absorbed by the above space bound. 
The boundaries between consecutive functions (the cumulative domain sizes \(\sum_{j=1}^{k}d_j\) and their cumulative range offsets \(\sum_{j=1}^{k}r_j\), for \(k\in[1\dd h]\)) are stored using EF encoding, requiring \(\cO(h\log(d/h)+h\log(r/h)+h)\) additional bits, which is absorbed by the above space bound. 

To answer a query \(f_k(i)\) specified by \((k, i)\), we first retrieve, in \(\cO(1)\) time, the cumulative domain boundary \(\sum_{j=1}^{k-1}d_j\), the cumulative range offset \(\sum_{j=1}^{k-1}r_j\) (both \(0\) if \(k=1\)), and \(Z_{k-1}\) and \(Z_k\). We then scan the stored breakpoints corresponding to the \(k\)-th function (i.e., those in positions \([Z_{k-1}+1\dd Z_k]\)) until finding the first breakpoint whose endpoint value, after subtracting the
cumulative domain offset, is at least \(i\). Finally, we return the corresponding stored value after removing the range offset. Suppose the scan finds the $j$-th breakpoint of the $k$-th function; the scan takes \(\cO(j)\) time. As in the proof of \Cref{lem:EF_small_range}, we have $j \le i$, and hence the query time is \(\cO(i)\).
\end{proof}

\subsection{Heavy-Path Decomposition}
  The \emph{heavy-path decomposition} of a rooted tree $\mathcal{T}$ is defined as follows~\cite{DBLP:journals/jcss/SleatorT83}. Each node of $\mathcal{T}$ is classified as either $\light$ or $\heavy$. The root is $\light$, and for every internal node, exactly one child is designated as \heavy; namely, the child with the largest subtree size (ties are broken arbitrarily), where the subtree size is defined as the number of leaves in the subtree. The edge from a node to its $\heavy$ child is called a \emph{heavy edge}, while all other child edges are called \emph{light edges}. An important property of this decomposition is that every root-to-leaf path contains at most $\log n$ light edges, where $n$ is the total number of leaves in $\mathcal{T}$. A \emph{heavy path} is a maximal path consisting entirely of heavy edges. 
  The heavy-path decomposition thus partitions $\mathcal{T}$ into a set of heavy paths.
  Since each node has at most one $\heavy$ child, every node belongs to exactly one heavy path. Each heavy path contains exactly one $\light$ node; namely, its topmost node. Starting from this $\light$ node, the remainder of the path is obtained by repeatedly following $\heavy$ children downward until a leaf node is reached. Therefore, each heavy path contains exactly one leaf node. We will assume throughout, without loss of generality, that each internal node has at least two children. Indeed, this is always the case for the class of suffix trees.

\begin{lemma}\label{lem:HP}
 A rooted ordered tree $\mathcal{T}$ with $n$ leaves, such that each internal node has at least two children, can be preprocessed into an $\cO(n)$-bit data structure that supports the following query in $\cO(1)$ time: given a node $u$, specified by its preorder rank, return the unique leaf on the same heavy path as $u$.   
\end{lemma}
\begin{proof}
 Preprocess the topology of $\mathcal{T}$ into an $\cO(n)$-bit data structure that supports standard navigational operations in $\cO(1)$ time~\cite[Table 1]{DBLP:journals/talg/NavarroS14}. Also, let $A[1\dd n]$ be an array, such that $A[i]$ denotes the node depth of the root of the heavy path containing $\ell_i$, the $i$-th leftmost leaf (e.g., output in a depth-first traversal).
 Here the depth of a node is defined as the number of its proper ancestors.
 We maintain a range minimum query (RMQ) data structure over $A$, which takes $\cO(n)$ bits and supports queries in $\cO(1)$ time~\cite{DBLP:journals/siamcomp/FischerH11}.
 Array $A$ is not stored.
 Finally, let $B[1 \dd n']$ be a bit vector, such that $B[i]=1$
 if and only if the node with preorder rank $i$ is a leaf of $\mathcal{T}$; here, $n'<2n$ denotes the total number of nodes in $\mathcal{T}$.
We maintain $B$ and augment it with an $\cO(1)$-time rank-select structure
~\cite{DBLP:conf/focs/Jacobson89}. The total space is $\cO(n)$ bits. 

Given a query node $u$, specified by its preorder rank, first identify the preorder rank $p_a$ of the leftmost leaf $\ell_a$ and the preorder rank $p_b$ of the rightmost leaf $\ell_b$ in the subtree rooted at $u$ in $\cO(1)$ time using the tree topology of $\mathcal{T}$. Next, use the bit vector $B$ to determine the range $[a \dd b]$, where $a$ (respectively, $b$) is the number of $1$'s in $B[1 \dd p_a]$ (respectively, $B[1 \dd p_b]$), computed via rank queries on $B$ in $\cO(1)$ time. Finally, perform an RMQ on $A$ over the range $[a \dd b]$. Suppose the answer returned is $k \in [a\dd b]$ (i.e., $A[k]$ is a smallest element in $A[a\dd b]$). We claim that $\ell_k$ is the unique leaf on the same heavy path as $u$.
 To see why, observe that among all leaves in the subtree of $u$, exactly one leaf lies on the same heavy path as $u$; namely, the leaf whose heavy path has the shallowest root. This is because:
\begin{itemize}
    \item If a leaf $\ell$ is in the subtree of $u$ and lies on the same heavy path as $u$, then the root of $\ell$'s heavy path is an ancestor of $u$ (possibly $u$ itself).
    \item If a leaf $\ell'$ is in the subtree of $u$ but lies on a different heavy path, then the root of $\ell'$'s heavy path is a proper descendant of $u$, and thus has greater depth than the root of $u$'s heavy path.
\end{itemize}
Therefore, the RMQ, which returns the leaf with minimum $A[k]$ (i.e., shallowest heavy-path root), correctly identifies the unique leaf on the same heavy path as $u$. This completes the proof.
 \end{proof}

\section{An $\cO(n\log\sigma +n\log^2\log n)$-bit and $\cO(|P|)$-time Solution}\label{sec:optimal_time}

We maintain the representation of $\SA_T$ and the topology of $\ST_T$ as described in \Cref{sec:prel}. 
Additionally, for each node of $\ST_T$, we maintain a bit indicating whether the node is \light or \heavy in the heavy-path decomposition of $\ST_T$. We will use the notation $\hpl(u)=k$ to denote that $\ell_k$ is the unique leaf on the same heavy path as node $u$ in $\ST_T$. To this end, we maintain the $\cO(n)$-bit structure of \Cref{lem:HP}, so that $\hpl(\cdot)$ can be evaluated in $\cO(1)$ time. 
All of these data structures occupy $\cO(n\log\sigma)$ bits.

For each \(k \in [1\dd n]\), we define the function \(f_k\) as follows
\[
f_k(i)\coloneqq\phi_T(\str(\ell_k)[1\dd i]);
\]
i.e., \(f_k(i)\) is the value of \(\phi_T\) on the prefix of length \(i\) along the path from the root of $\ST_T$ to leaf \(\ell_k\). For \(i>|\str(\ell_k)|\), we set \(f_k(i)\coloneqq 0\).

Therefore, \Cref{prob:main} boils down to encoding these functions efficiently, such that, given a pattern \(P\), we can return \(f_q(|P|)\), where \(\hpl(\locus(P))=q\), if the locus of $P$ exists in $\ST_T$. 

We now present the details of the data structures associated with  $\ell_k$, which enable efficient processing of queries $P$ satisfying \(\hpl(\locus(P))=k\).
Let \(w_k\) be the lowest \light ancestor of \(\ell_k\); this is the root of the heavy path ending with \(\ell_k\). (If $w_k$ is the root of $\ST_T$, we set $|\str(\parent(w_k))|\coloneqq 0$.)
We aim to identify a node $u_k$ on the path from $w_k$ to $\ell_k$. 

\begin{itemize}
    \item {\bf Reporting Case:}   If $|\str(\parent(w_k))| > \size(w_k) \cdot \log n$, then for any query pattern $P$ satisfying $\hpl(\locus(P))=k$, we have $|P| > |\Occ_T(P)| \cdot \log n$. Therefore, such queries can be answered by first reporting $\Occ_T(P)$ (using the $\SA_T$ representation), and then using Condition 1 of \Cref{prob:main}, for a total query time of $\cO(|P|)$. There is no need to store the function \(f_k\) in this case; we set \(u_k \coloneqq w_k\). We mark the leaves \(\ell_k\) for which \(f_k\) is stored in a bit vector of \(n\) bits with rank-select support~\cite{DBLP:conf/focs/Jacobson89}, so that the index of \(f_k\) among the stored functions is obtained in \(\cO(1)\) time.

    \item \noindent{\bf Lookup Case:} If $|\str(\parent(w_k))| \leq \size(w_k) \cdot \log n$, then there may exist query patterns \(P\) such that \(\hpl({\locus}(P))=k\) and \(|P| \leq |\Occ_T(P)| \cdot \log n\). Therefore, we cannot afford to report all occurrences (and process them) in this case. We split this case into two subcases as follows: 
\begin{enumerate}
    \item {\bf Subcase 1: \(|\str(\parent(w_k))| \cdot \log n \leq \size(w_k)\).}
We define \(u_k\) to be the deepest node (i.e., the node farthest from the root) on the path from \(w_k\) to \(\ell_k\) satisfying
\[|\str(\parent(u_k))|\cdot \log n \leq \size(u_k).\] 
Then, we explicitly store \(f_k(i)\), for all \(i \in [1\dd |\str(\parent(u_k))|]\), as a list \(L_k\) supporting \(\cO(1)\)-time lookup. By the above inequality, the space usage is \(\frac{\size(u_k)}{\log n}\log n=\cO(\size(u_k))\) bits. To store \(f_k(i)\), for all \(i \in [|\str(\parent(u_k))|+1\dd \size(u_k)\log n]\), we use EF encoding. First, note that the right endpoint of the above interval is sufficient as otherwise we are in the reporting case. Second, during the encoding, we treat \(f_k(i)\), for all \(i \leq |\str(\parent(u_k))|\), as \(\size(u_k)\) to preserve monotonicity (Condition 2 of \Cref{prob:main}). The key observation is that \(f_k(i) \in [0\dd\size(u_k)]\); i.e., the domain size is \(\cO(\size(u_k)\log n)\), while the range size is \(\cO(\size(u_k))\). 
Therefore, we add \(f_k\) to the collection of functions encoded using Lemma~\ref{lem:EF_opt_time}.

\item {\bf Subcase 2: \(|\str(\parent(w_k))| \cdot \log n > \size(w_k)\).}
We simply set \(u_k \coloneqq w_k\). The list \(L_k\) is empty in this subcase, and \(f_k(i)\), for all \(i \in [|\str(\parent(u_k))|+1\dd \size(u_k)\log n]\), is encoded using EF encoding as in the previous subcase. As observed earlier, \(f_k(i) \in [0\dd\size(u_k)]\); i.e., the domain size is \(\cO(\size(u_k)\log n)\), while the range size is \(\cO(\size(u_k))\). Therefore, we add \(f_k\) to the collection of functions encoded using Lemma~\ref{lem:EF_opt_time}.
\end{enumerate}
The total query time is \(\cO(|P|)\) (in any case): finding the locus of $P$ takes \(\cO(|P|)\) time; table lookup takes \(\cO(1)\) time; and decoding a value from the structure in \Cref{lem:EF_opt_time} takes \(\cO(|P|)\) time.
\end{itemize}

Let us now bound the space complexity. 
First, we concatenate all non-empty lists \(L_k\) in increasing order of \(k\) to obtain a single list \(L_T\). Additionally, we maintain a bit vector \(B_T \coloneqq 10^{|L_1|}10^{|L_2|}10^{|L_3|}\dots 10^{|L_n|}\), where \(10^{|L_k|}\) denotes a $1$ followed by \(|L_k|\) zeros (and if \(|L_k|=0\), this is just a 1). We then augment \(B_T\) with an \(\cO(1)\)-time rank-select structure~\cite{DBLP:conf/focs/Jacobson89}---the total space is \(\cO(n+|L_T|)\) bits. Now, a query \(L_k[i]\), for a non-empty list \(L_k\), is answered by returning \(L_T[j+i]\), where \(j\) is the number of zeros preceding the \(k\)-th $1$ in \(B_T\), which can be computed in \(\cO(1)\) time via the rank-select structure. The space (in bits) taken by \(L_T\), i.e., 
\(\cO(|L_T|\log n)\), 
is linear in the sum of \(\size(u_k)\) over all \(u_k\) that fall under Subcase 1, which is \(\cO(n)\), since there is no ancestor--descendant relationship between any two such \(u_k\)'s. Indeed, suppose for contradiction that $u_{k'}$ is a proper ancestor of $u_k$, for two such nodes $u_k$ and $u_{k'}$. Let $h$ be the \heavy child of $u_{k'}$; it lies on the heavy path of $\ell_{k'}$, strictly below $u_{k'}$. Since $u_k$ lies in the subtree of some child of $u_{k'}$, and $h$ has the largest subtree among the children of $u_{k'}$, we have $\size(h)\geq\size(u_k)$. Moreover, $\parent(u_k)$ is a descendant of $u_{k'}$ (possibly $u_{k'}$ itself), so $|\str(u_{k'})|\leq|\str(\parent(u_k))|$. Hence, $|\str(\parent(h))|\cdot \log n=|\str(u_{k'})|\cdot \log n\leq|\str(\parent(u_k))|\cdot \log n\leq\size(u_k)\leq\size(h)$; that is, $h$ also satisfies the inequality defining $u_{k'}$, contradicting the choice of $u_{k'}$ as the deepest such node on its heavy path.
Therefore, the combined size of $B_T$ and $L_T$ is $\cO(n)$ bits.

Next, we bound the space of all \(f_k\)'s that are encoded using Lemma~\ref{lem:EF_opt_time}. For Subcase 1, the $u_k$'s are disjoint (no ancestor--descendant relationships), so each leaf is in the subtree of at most one such $u_k$.
For Subcase 2, we need the following simple lemma.

\begin{lemma}\label{lem:loglog}
    Let $n'$ and $n$ be two positive integers with $n' \leq n$. Starting from $x_0 \coloneqq n' \log n$ and repeatedly dividing by $2$ (i.e., $x_{t+1} = x_t / 2$), the number of divisions required until $x_t < n' / \log n$ is $\cO(\log \log n)$.
\end{lemma}
\begin{proof}
Let $x_0 \coloneqq n' \log n$. After $t$ divisions by $2$, we have $x_t = \frac{n' \log n}{2^t}$. We want to find the smallest $t$ such that $x_t < n' / \log n$. This is $t = \lfloor 2 \log \log n \rfloor + 1 = \cO(\log \log n)$, which completes the proof.
\end{proof}

\begin{figure}[t]
    \centering
    \includegraphics[width=0.75\linewidth]{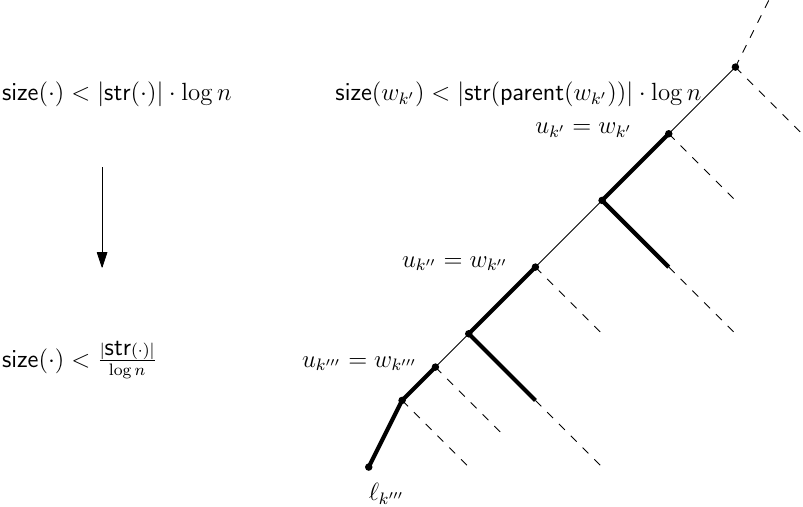}
    \caption{After visiting $\cO(\log\log n)$ \light nodes, as we traverse down a root-to-leaf path, the subtree size becomes sufficiently small, $|\str(\parent(w_{k'''}))| > \size(w_{k'''}) \cdot \log n$, that we can use reporting.}
    \label{fig:tree}
\end{figure}

Consider all \(u_k\)'s that fall under Subcase 2: they are all \light nodes because \(u_k=w_k\), and \(w_k\) is the root of a heavy path, which is always a \light node. The \(u_k\)'s may have ancestor--descendant relationships. However, we argue that the number of such \(u_k\)'s on any root-to-leaf path is bounded by \(\cO(\log\log n)\). This is because, as we traverse downwards, the subtree size decreases by at least a factor of two whenever we reach a \light node. Therefore, by \Cref{lem:loglog}, after visiting \(\cO(\log\log n)\) \light nodes, the subtree size becomes small enough that we can afford to report all occurrences. This means that any leaf of $\ST_T$ can be in the subtree of \(\cO(\log\log n)\) such \(u_k\)'s under Subcase 2. Inspect \Cref{fig:tree}. 

Therefore, for all functions $f_k$ stored using \Cref{lem:EF_opt_time} (in both Subcase 1 and Subcase 2), each function has range size $\cO(\size(u_k))$ and domain size $\cO(\size(u_k)\log n)$. 
Since each leaf belongs to the subtree of $\cO(\log\log n)$ nodes $u_k$, the sum of the sizes of all ranges is $\cO(n\log\log n)$, and the sum of the sizes of all domains is $\cO(n\log n\log\log n)$. 
Applying Lemma~\ref{lem:EF_opt_time} to these functions, we obtain a total space (in bits) of
\[
\cO\left(n\log\log n  \log\left(\frac{n \log n  \log\log n}{n\log\log n}\right)\right)
= \cO(n\log^2\log n).
\]

Adding the $\cO(n\log\sigma)$ bits of the other structures, we arrive at the claimed trade-off.

\section{An $\cO(n\log\sigma)$-bit and $\cO(|P|\log_{\sigma}^{\epsilon} n)$-time Solution}\label{sec:optimal_space}

Let \(\tau\geq 1\) be a parameter to be fixed later.
We modify the construction in \Cref{sec:optimal_time} as follows. Previously, in Subcases 1 and 2, we used EF encoding (\Cref{lem:EF_opt_time}) to store \(f_k(i)\), for \(i\leq \size(u_k)\log n\). Here, we instead store \(f_k(i)\), for \(i \leq \size(u_k)/\tau\). (If \(\lfloor\size(u_k)/\tau\rfloor=0\), we store nothing.) The sum of the sizes of all ranges remains \(\cO(n\log\log n)\), while the sum of the sizes of all domains becomes \(\cO((n/\tau)\log\log n)\). We set 
$\tau \coloneqq \log^2\log n$ and use \Cref{lem:EF_compact} instead, as the domain size is now small. The space for encoding all functions becomes:
\[
\cO\left(\frac{n \log \log n}{\tau} \cdot \log \tau\right) \subseteq \cO(n) \text{ bits}.
\]

The query time for the lookup case is now \(\cO(1)\) (see the contrast between \Cref{lem:EF_opt_time} and \Cref{lem:EF_compact}). However, for the reporting case, we need to account for the additional factor \(\tau\). In the reporting case, we have $|P| > \size(u_k)/\tau$, so $|\Occ_T(P)| \leq \size(u_k) < \tau |P|$. Therefore, the query time becomes:
\[
|P| + |\Occ_T(P)|\log\log n + |\Occ_T(P)|\log_{\sigma}^{\epsilon} n
\leq |P|\tau(\log\log n+\log_{\sigma}^{\epsilon} n)
= \cO(|P|\log^{2\epsilon} n).
\]

Finally, if $\log \sigma < \sqrt{\log n}$, then
\(
\log n=\log_\sigma n\cdot\log\sigma
    \le \log_\sigma n\cdot\log^{1/2}n,
\)
which implies
\[
\log^{1/2}n\le \log_\sigma n.
\]
Hence
\(
\log^{2\epsilon} n
    =(\log^{1/2}n)^{4\epsilon}
    \le \log_\sigma^{4\epsilon}n,
\)
and therefore
\(
\cO(|P|\log^{2\epsilon} n)
    =\cO(|P|\log_\sigma^{4\epsilon}n).
\)
Thus the query time can be written as
$\cO(|P|\log_\sigma^{\epsilon'}n)$, for the constant
$\epsilon'=4\epsilon$. Otherwise, when $\log \sigma \geq \sqrt{\log n}$, we use the construction from \Cref{sec:optimal_time} that now has space $\cO(n \log\sigma)$ bits and $\cO(|P|)$ query time.

We have arrived at \Cref{thm:main}.

\section{Applications}\label{sec:apps}

In this section, we showcase three headline applications of \Cref{thm:main} that improve upon the state-of-the-art indexes; however, we anticipate that \Cref{thm:main} has many more applications.

\subsection{Counting Non-Overlapping Occurrences}
\label{sec:non-overlap} 

Given a string~$S$ and two fragments of $S$, $S[i\dd i']$ and $S[j\dd j']$ with $i \leq j$, we say that the fragments are \emph{non-overlapping} if $j > i'$.
The set of non-overlapping occurrences of a string $P$ in a string $T$ is a \emph{largest} set of occurrences of $P$ in $T$ that are pairwise non-overlapping.
We denote this set by $\Occno_T(P)$. 

\defDSproblem{Non-Overlapping Occurrences Counting (\NOC)}
{A text $T=T[1\dd n]\in[0\dd \sigma)^n$.
}
{Given a pattern $P$ of length $m$, return $|\Occno_T(P)|$.}

Reporting or counting non-overlapping occurrences is a well-studied problem~\cite{DBLP:journals/algorithmica/ApostolicoP96,DBLP:conf/icalp/BrodalLOP02,DBLP:conf/isaac/CohenP09,DBLP:journals/algorithmica/GangulyST20,DBLP:journals/tcs/HooshmandAKT21,DBLP:journals/tcs/GibneyMT25,DBLP:conf/esa/Bals26}. The following classic result is known for \NOC.

\begin{proposition}[\cite{DBLP:journals/algorithmica/ApostolicoP96,DBLP:conf/icalp/BrodalLOP02}]\label{the:noo-sota1}
    There exists an $\cO(n\log n)$-bit data structure for \NOC with $\cO(m)$ query time.
\end{proposition}

Bals et al.~\cite{DBLP:conf/esa/Bals26} also showed a different trade-off for \NOC.

\begin{proposition}[\cite{DBLP:conf/esa/Bals26}]\label{the:noo-sota2}
    There exists an $\cO(n\log \sigma)$-bit data structure for \NOC with $\cO(m\log n \log_{\sigma}^{\epsilon} n)$ query time.
\end{proposition}

We improve upon both results as follows.

\begin{theorem}\label{the:noo}
    There exists an $\cO(n \log \sigma + n(\log \log n)^2)$-bit data structure for \NOC with $\cO(m)$ query time.
    Alternatively, there exists an $\cO(n \log \sigma)$-bit data structure supporting \NOC queries in $\cO(m\log_{\sigma}^{\epsilon} n)$ time.
\end{theorem}
\begin{proof}
    We simply apply \Cref{thm:main} setting $\phi_T(P)\coloneqq|\Occno_T(P)|$. Then it is easy to verify that: 
    \begin{itemize}
        \item \(\phi_T(P)\in [0\dd |\Occ_T(P)|]\), and that it can be computed from the set \(\Occ_T(P)\) of occurrences of \(P\) in \(T\) in \(\cO(|\Occ_T(P)| \log\log n)\) time, via integer sorting~\cite{DBLP:journals/jal/Han04} and a linear scan, as desired in \Cref{prob:main}.
        \item $\phi_T(P)=|\Occno_T(P)|$ is monotone as desired in \Cref{prob:main}. Let $Q$ be a prefix of $P$.
        Then extending $Q$ to $P$ both reduces the available occurrences \emph{and} increases the gap requirement between the occurrences of $P$. Consequently, $\phi_T(P) \leq \phi_T(Q)$.
    \end{itemize}
    This completes the proof.
\end{proof}

\subsection{Counting Consecutive Occurrences}
\label{sec:consecutive} 

Given a string $T$, an integer $\beta>0$, and a pattern $P$, a pair $(i,j)$, with $i<j$, is called a $\beta$-\emph{consecutive occurrence} of $P$ in $T$ if $i,j\in \Occ_T(P)$, there is no other occurrence of $P$ between $i$ and $j$, and $(j - i) \leq \beta$. We denote this set of pairs by $\Occco_T(P)$.

\defDSproblem{Consecutive Occurrences Counting (\COC)}
{A text $T=T[1\dd n]\in[0\dd \sigma)^n$ and an integer $\beta>0$.}
{Given a pattern $P$ of length $m$, return $|\Occco_T(P)|$.}

Many variants of consecutive occurrences have been studied in the literature for both reporting and counting~\cite{DBLP:journals/algorithmica/BilleG14,
DBLP:conf/stacs/BilleGLPRS24,
DBLP:journals/tcs/BilleGPRS22,
DBLP:journals/algorithmica/BilleGPS23,
DBLP:journals/mst/BilleGVV14,
DBLP:conf/cpm/BrodalLPS99,
DBLP:journals/is/GawrychowskiGSS26,
DBLP:journals/algorithmica/IliopoulosR09,
DBLP:conf/wads/KellerKL07,
DBLP:conf/soda/Muthukrishnan02,
DBLP:journals/tcs/NavarroT16,
DBLP:conf/esa/Radoszewski23,DBLP:conf/esa/Bals26}. 
Bals et al.~\cite{DBLP:conf/esa/Bals26} showed the following result for \COC.

\begin{proposition}[\cite{DBLP:conf/esa/Bals26}]\label{the:coc-sota}
    There exists an $\cO(n\log n)$-bit data structure for \COC with $\cO(m)$ query time.
\end{proposition}

We improve upon this result as follows.

\begin{theorem}\label{the:coc}
    There exists an $\cO(n \log \sigma + n(\log \log n)^2)$-bit data structure for \COC with $\cO(m)$ query time.
       Alternatively, there exists an $\cO(n \log \sigma)$-bit data structure supporting \COC queries in $\cO(m\log_{\sigma}^{\epsilon} n)$ time.
\end{theorem}
\begin{proof}
    We simply apply \Cref{thm:main} setting $\phi_T(P)\coloneqq|\Occco_T(P)|$. Then it is easy to verify that: 
    \begin{itemize}
        \item \(\phi_T(P)\in [0\dd |\Occ_T(P)|]\), and that it can be computed from the set \(\Occ_T(P)\) of occurrences of \(P\) in \(T\) in \(\cO(|\Occ_T(P)| \log\log n)\) time, via integer sorting~\cite{DBLP:journals/jal/Han04} and a linear scan, as desired in \Cref{prob:main}.
        \item $\phi_T(P)=|\Occco_T(P)|$ is monotone as desired in \Cref{prob:main}. Let $Q$ be a prefix of $P$.
        Then extending $Q$ to $P$ both reduces the available occurrences \emph{and} makes the gap of consecutive occurrences of $P$ larger (or the same).  Consequently, $\phi_T(P) \leq \phi_T(Q)$.
    \end{itemize}
    This completes the proof.
\end{proof}

\subsection{Counting Property-Restricted Occurrences}
\label{sec:properties} 

A \emph{property} $\Pi$ of a text $T$ of length $n$ is a hereditary collection of integer intervals contained in $\{1, \ldots, n\}$ (hereditary means closed under subintervals). For simplicity, we represent every property $\Pi$ with an array $\pi[1 \dd n]$ such that the longest interval $I \in \Pi$ starting at position $i$ is $\{i, \ldots, \pi[i]\}$. Observe that $\pi$ can be an arbitrary array satisfying $\pi[i] \in \{i-1, \ldots, n\}$ and $\pi[1] \leq \pi[2] \leq \cdots \leq \pi[n]$. For a given pattern $P$, by $\Occ_T^\pi(P)$ we denote the set of occurrences $i$ of $P$ in $T$ such that $i + |P| - 1 \leq \pi[i]$. 

\defDSproblem{Property-Restricted Occurrences Counting (\POC)}
{A text $T=T[1\dd n]\in[0\dd \sigma)^n$ and an array $\pi[1\dd n]$ representing a hereditary property $\Pi$.}
{Given a pattern $P$ of length $m$, return $|\Occ_T^\pi(P)|$.}

Reporting or counting property-restricted occurrences is a well-studied problem~\cite{DBLP:journals/tcs/AmirCIKZ08,DBLP:journals/ipl/IliopoulosR08,DBLP:journals/ipl/JuanLW09,DBLP:journals/iandc/HonPST13,DBLP:journals/iandc/BartonK0PR20,DBLP:journals/jea/Charalampopoulos20,DBLP:conf/esa/Bals26}. The following result is known in particular for \POC.

\begin{proposition}[\cite{DBLP:journals/iandc/BartonK0PR20}]\label{the:prc-sota}
    There exists an $\cO(n\log n)$-bit data structure for \POC with $\cO(m)$ query time.
\end{proposition}

We improve upon this result as follows.

\begin{theorem}\label{the:prc}
    There exists an $\cO(n \log \sigma + n(\log \log n)^2)$-bit data structure for \POC with $\cO(m)$ query time.
       Alternatively, there exists an $\cO(n \log \sigma)$-bit data structure supporting \POC queries in $\cO(m\log_{\sigma}^{\epsilon} n)$ time.
\end{theorem}
\begin{proof}

The array $\pi$ can be encoded using EF encoding in $\cO(n)$ bits, while supporting $\cO(1)$-time retrieval of $\pi[i]$, for any given index $i$.

Then, we simply apply \Cref{thm:main} setting $\phi_T(P)\coloneqq|\Occ_T^\pi(P)|$. It is easy to verify that: 
    \begin{itemize}
        \item \(\phi_T(P)\in [0\dd |\Occ_T(P)|]\), and that it can be computed from the set \(\Occ_T(P)\) of occurrences of \(P\) in \(T\) in \(\cO(|\Occ_T(P)|)\) time, via $\pi[\cdot]$ queries,
        as desired in \Cref{prob:main}.
        \item $\phi_T(P)=|\Occ_T^\pi(P)|$ is monotone as desired in \Cref{prob:main}. Let $Q$ be a prefix of $P$. Then $\Occ_T(P) \subseteq \Occ_T(Q)$. Moreover, for each $i \in \Occ_T(P)$, $i + |P| - 1 \leq \pi[i]$ implies $i + |Q| - 1 \leq i + |P| - 1 \leq \pi[i]$, so every occurrence counted for $P$ is also counted for $Q$. Thus, $\phi_T(P) \leq \phi_T(Q)$.
    \end{itemize}
    This completes the proof.
\end{proof}

\section*{Declarations}
All scientific ideas and results in this paper are due to the authors, who wrote the manuscript. Large language models were subsequently used to proofread it: they flagged minor errors and inconsistencies and suggested corrections and stylistic improvements. The authors independently verified every suggestion and decided whether and how to incorporate it. The authors take full responsibility for the entire content.

\bibliographystyle{alphaurl}
\bibliography{references}

\end{document}